\documentclass{pasj01}

\usepackage{natbib}
\usepackage{color}
\Received{$\langle$reception date$\rangle$}
\Accepted{$\langle$acception date$\rangle$}
\Published{$\langle$publication date$\rangle$}

\begin{document}

\title{ Do gas clouds in narrow-line regions of Seyfert galaxies come from their nuclei? }

\author{Kazuma Joh${}^1$, Tohru Nagao${}^2$, Keiichi Wada${}^{3,2}$, Koki Terao${}^4$, and Takuji Yamashita${}^{5,2}$}%
\altaffiltext{}{${}^1$Graduate School of Science and Engineering, Ehime University, Bunkyo-cho, Matsuyama 790-8577, Japan \\
${}^2$Research Center for Space and Cosmic Evolution, Ehime University, Bunkyo-cho, Matsuyama 790-8577, Japan \\
${}^3$Kagoshima University, Graduate School of Science and Engineering, Kagoshima 890-0065, Japan \\
${}^4$Astronomical Institute, Tohoku University, 6-3 Aramaki, Aoba-ku, Sendai 980-8578, Japan \\
${}^5$National Astronomical Observatory of Japan, 2-21-1 Osawa, Mitaka, Tokyo 181-8588, Japan}

\email{joh@cosmos.phys.sci.ehime-u.ac.jp}


\maketitle

\begin{abstract}
The narrow-line region (NLR) consists of gas clouds ionized by the strong radiation from the active galactic nucleus (AGN), distributed in the spatial scale of AGN host galaxies. The strong emission lines from the NLR are useful to diagnose physical and chemical properties of the interstellar medium in AGN host galaxies. However, the origin of the NLR is unclear; the gas clouds in NLRs may be originally in the host and photoionized by the AGN radiation, or they may be transferred from the nucleus with AGN-driven outflows. For studying the origin of the NLR, we systematically investigate the gas density and velocity dispersion of NLR gas clouds using a large spectroscopic data set taken from the Sloan Digital Sky Survey. The [S~{\sc ii}] emission-line flux ratio and [O~{\sc iii}] velocity width of 9,571 type-2 Seyfert galaxies and 110,041 star-forming galaxies suggest that the gas density and velocity dispersion of NLR clouds in Seyfert galaxies ($n_{\rm e} \approx$ 194 cm$^{-3}$ and $\sigma_{\rm [O~{\sc III}]} \approx$ 147 km s$^{-1}$) are systematically larger than those of clouds in H~{\sc ii} regions of star-forming galaxies ($n_{\rm e} \approx$ 29 cm$^{-3}$ and $\sigma_{\rm [O~{\sc III}]} \approx$ 58 km s$^{-1}$). Interestingly, the electron density and velocity dispersion of NLR gas clouds are larger for Seyfert galaxies with a higher [O~{\sc iii}]/H$\beta$ flux ratio, i.e., with a more active AGN. We also investigate the spatially-resolved kinematics of ionized gas clouds using the Mapping Nearby Galaxies at the Apache Point Observatory (MaNGA) survey data for 90 Seyfert galaxies and 801 star-forming galaxies. We find that the velocity dispersion of NLR gas clouds in Seyfert galaxies is larger than that in star-forming galaxies at a fixed stellar mass, at both central and off-central regions. These results suggest that gas clouds in NLRs come from the nucleus, probably through AGN outflows.
\end{abstract}

\section{Introduction}
There is accumulating evidence of the so-called co-evolution \citep[e.g.,][]{Kor13} of galaxies and super massive black holes (SMBHs), since the correlation between masses of SMBHs and properties of their host galaxies has been reported observationally \citep[e.g.,][]{Fer00, Geb00, McC13, Rei15}. For studying the physics of the co-evolution, active galactic nuclei (AGNs) are important because SMBHs in AGNs are in a growing phase due to the mass accretion on to the SMBH. AGNs are also important to understand the growth of their host galaxies, because the AGN activity sometimes affects the ISM in their host galaxies. Specifically, AGN driven outflows have been detected at the spatial scale of AGN host galaxies by observations of multi-phase gas \citep[e.g.,][]{Vil11,Cic14, Kan18}. Such AGN-driven outflows may contribute significantly to the suppression of the star formation in the host galaxy, and accordingly the control of mass accretion to the SMBH \citep[e.g.,][]{Can12, Hec14, For19}, while the AGN-driven outflows may enhance the star formation due to the compression of the gas \citep[e.g.,][]{Mai17, Shi19, Zhu20}.

To investigate the co-evolution observationally, the narrow-line region (NLR) is a useful ingredient in the AGN. This is because the NLR in nearby AGNs is spatially resolved observationally (with the typical extension of $\sim10^{2-4}$ pc) and it shows strong emission lines in a wide wavelength range. Therefore the NLR is powerful to study the spatial distribution of the physical, chemical, and kinematic properties of gas in AGN host galaxies. The mechanism of the ionization of NLR gas clouds is well understood; they are mostly ionized by the power-law radiation from the AGN \citep[e.g.,][]{Dav79, Eva86, Bin96, Kom97, Nag06}, though sometimes fast shocks may also contribute \citep[e.g.,][]{Wil99, Fu07, Has11, Ter16}. On the other hand, the origin of gas clouds in NLRs is not clearly identified. One possible scenario is as follows; gas clouds originally located in the host galaxy become the NLR when they are ionized by the power-law radiation from the AGN. Another idea is that the NLR clouds are carried in by the AGN-driven outflows from the central region. \citet{Wad18} showed that the NLR can be explained by high-density gas clouds transferred from the nuclear region to the outer part in the host galaxy through the AGN-driven outflow, based on three-dimensional radiation-hydrodynamical calculations \citep[see also][]{Wad12}. Although the spatial scale of the NLR simulated by \citet{Wad18} was limited to a region of $\sim$ 10 pc from the nucleus, a more extended NLR can be formed through the same scenario since the velocity of these gas clouds is larger than the escape velocity. In fact, the gas in the NLR are located mostly at the inner surface of the bi-cone in nearby AGNs \citep[e.g.,][]{Das06, Mul11}. This is also the case in the theoretical model.

These two possible scenarios for the origin of the NLR clouds can be distinguished by the statistical properties of the gas density and kinematics of NLR gas clouds, because the latter scenario predicts systematically higher gas density and more disturbed kinematics of NLR gas clouds than the interstellar medium (ISM) in non-AGN galaxies. \citet{Zha13} conducted a statistical study of the gas clouds in NLRs and reported that the typical range of the gas density (traced by the electron density) in NLRs is $\sim10^{2-3}~\rm{cm^{-3}}$, which is higher than that in H~{\sc ii} regions of star-forming galaxies ($\sim10^{1-2}~\rm{cm^{-3}}$). \citet{Zha13} investigated also the kinematics of NLR gas clouds, and reported that the velocity dispersion and gas density of NLR clouds are not correlated significantly \citep[see also][]{Hec81}. This seems to be contrary to the latter scenario of the origin of the NLR described above. 

The success of the Sloan Digital Sky Survey \citep[SDSS;][]{Yor00} has enabled us to examine the large spectroscopic data set to systematically investigate the physical and kinematic properties of emission-line galaxies such as AGNs and star-forming galaxies. In addition, the recent development of integral field unit (IFU) spectrographs has allowed us to examine the spatially resolved properties of galaxies. For instance, the Mapping Nearby Galaxies at the Apache Point Observatory \citep[MaNGA;][]{Bun15} is one of spatially-resolved spectroscopic surveys for nearby galaxies, which has made a significant contribution to the understanding of the galaxy evolution \citep[e.g.,][]{Bel16, Bel17, God17}. IFU observations have played important role also in investigating AGN-driven outflows in detail \citep[e.g.][]{Dav16, Min19}. Moreover, the development of IFU observations has led to an increase in the number of samples, which allows us to statistically study the effects of AGNs for the host galaxies \citep[e.g.,][]{Ilh19, Rod19}. 

In this work, we analyze the SDSS and MaNGA datasets to understand the origin of NLR gas clouds.
The paper is structured as follows. In section 2 we introduce the galaxy sample and the data analysis, and explain our approach to investigate the physical properties and ionization states of our emission-line galaxies. The results are described in section 3, and based on that, in section 4 the origin of the gas clouds in NLRs. We summarize the main results in section 5.

\section{Data and Analysis}

\subsection{The Sloan Digital Sky Survey (SDSS) data}

\subsubsection{Sample selection and classification}

We use a spectroscopic data set of nearby emission-line galaxies taken from the SDSS Data Release 8 \citep{Aih11}, which covers about a quarter of the entire sky. Those spectra were obtained with 3$^{\prime\prime}$-diameter fibers, thus there is no spatially-resolved spectral information in the data. The emission-line properties of SDSS emission-line galaxies such as fluxes and velocity dispersions were measured and provided as the Max Planck Institute for Astrophysics and Johns Hopkins University (MPA-JHU) catalog \citep{Kau03a, Bri04, Tre04}. To exclude duplicated objects in this catalog, we select objects with {\tt sciencePrimary} = 1 from the MPA-JHU catalog. This {\tt sciencePrimary} flag in the {\tt specObjAll} table is designed to select the best available unique set of spectra. After resolving the object duplications, we create a ``clean sample'' by selecting objects whose measurements are reliable by the following criteria. First, we exclude objects with {\tt RELIABLE} = 0, since spectroscopic parameters of such objects were not reliably measured by the SDSS pipeline. We also require that the redshift measurement was successful (i.e., {\tt Z\_WARNING} = 0) and $z>0.02$. This redshift limit is adopted to exclude too nearby galaxies, for which the SDSS fiber covers only very narrow area around the center of galaxies. We then remove type-1 AGNs because we cannot measure some key emission-line flux ratios such as [N~{\sc ii}]$\lambda$6584/H$\alpha$ and [O~{\sc iii}]$\lambda$5007/H$\beta$ due to the broad component of recombination lines. For removing them, we select only objects whose velocity width ($\sigma$) of both forbidden lines and Balmer lines is less than 500 km s$^{-1}$. Then we require signal-to-noise ratio (S/N) $>$ 7 for some key emission lines ($\rm{H\alpha}$, $\rm{H\beta}$, $\rm{[O\ {\scriptstyle III}]}\lambda5007$, $\rm{[N\ {\scriptstyle II}]}\lambda6584$ and $\rm{[S\ {\scriptstyle II}]}\lambda\lambda6717,6731$) and S/N $>$ 3 for $\rm{[O\ {\scriptstyle I}]}\lambda6300$ (since this emission line is the weakest among the diagnostic emission lines used in this work) for accurate calculations of their flux ratios. As the result of the above selection criteria, we obtain the clean sample that consists of 140,320 emission-line galaxies.

\begin{figure}
\begin{center}
\includegraphics[width=84mm]{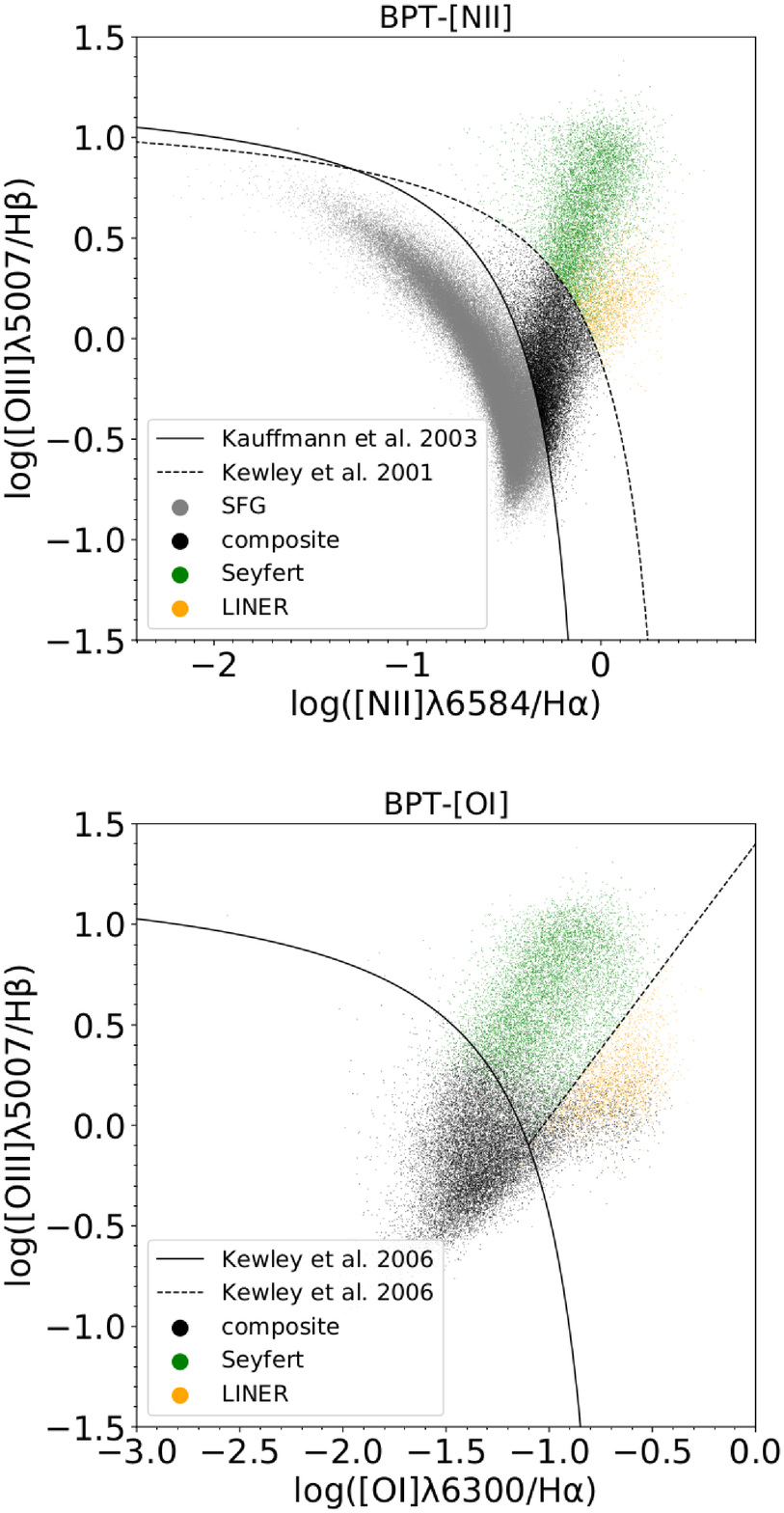}
\end{center}
\caption{BPT diagrams (upper panel: $\rm{[O\ {\scriptstyle III}]\lambda5007/H\beta}$ vs $\rm{[N\ {\scriptstyle II}]\lambda6584/H\alpha}$, lower panel: $\rm{[O\ {\scriptstyle III}]\lambda5007/H\beta}$ vs $\rm{[O\ {\scriptstyle I}]\lambda6300/H\alpha}$ vs. $\rm{[O\ {\scriptstyle III}]\lambda5007/H\beta}$), showing the classification criteria for SDSS emission-line galaxies. The solid and dotted lines in the top panel show the criteria selecting star-forming galaxies \citet{Kau03b} and composite galaxies \citet{Kew01}, respectively. The dotted line \citep{Kew06} in the lower panel separates the remaining galaxies into Seyfert galaxies and LINERs. Gray, black, green, and yellow dots demote star-forming galaxies, composite galaxies, Seyfert galaxies, and LINERs, respectively.}
\label{bpt}
\end{figure}

Next, we classify these 140,320 galaxies using two ionization diagnostic diagrams \citep[so-called BPT diagrams;][]{Bal81}, which use two pairs of emission-line flux ratios to classify emission-line galaxies into Seyfert galaxies, low-ionization nuclear emission-line regions (LINERs), star-forming galaxies, and composite galaxies (figure \ref{bpt}). More specifically, we adopt the following criteria for this classification.

\begin{enumerate}
\item The empirical classification criterion derived by \citet{Kau03b};
    \begin{eqnarray}
    \log \left( \frac{\rm [O\ {\scriptstyle III}]}{\rm H\beta} \right) > \frac{0.61}{\rm log ([N\ {\scriptstyle II}]/H\alpha) -0.05}+1.3,
    \end{eqnarray}
    which separates star-forming galaxies (110,041 objects) from the remaining galaxies.
\item The theoretical boundary of the region where star-forming galaxies can distribute on the BPT diagram, which derived by \citet{Kew01} by combining stellar population synthesis models and photoionization models,
    \begin{eqnarray}
    \log \left( \frac{\rm [O\ {\scriptstyle III}]}{\rm H\beta} \right) > \frac{0.61}{\rm log ([N\ {\scriptstyle II}]/H\alpha) -0.47}+1.19,
    \end{eqnarray}
    which defines composite galaxies (18,523 objects).
\item The empirical classification criterion derived by \citet{Kew06},
    \begin{eqnarray}
    \log \left( \frac{\rm [O\ {\scriptstyle III}]}{\rm H\beta} \right) > 1.36\log \left( \frac{\rm [O\ {\scriptstyle I}]}{\rm H\alpha} \right) + 1.4, 
    \end{eqnarray}
    which classifies the remaining objects into Seyfert galaxies (9,571 objects) and LINERs (2,185 objects).
\end{enumerate}

Figure \ref{bpt} shows the BPT diagrams with the result of our classification of the SDSS clean sample.
The averages and standard deviations of the redshift of the selected star-forming galaxies, composite galaxies, Seyfert galaxies, and LINERs are $0.077 \pm 0.044$, $0.095 \pm 0.049$, $0.104 \pm 0.048$, and $0.079 \pm 0.047$, respectively.

\subsubsection{Electron density}

In order to estimate the gas density of ionized clouds, the electron density inferred by the emission line ratio of $\rm{[S\ {\scriptstyle II}]}\lambda\lambda6717,6731$ has been often used \citep[e.g.,][]{Ost06}. To derive the electron density, we use the equation shown by \citet{San16},
\begin{equation}
 	n_{\rm{e}}=\frac{cR_{\rm[S\ {\scriptstyle II}]}-ab}{a-R_{\rm[S\ {\scriptstyle II}]}},
\end{equation}
where $R_{\rm[S\ {\scriptstyle II}]} = \rm{[S\ {\scriptstyle II}]}\lambda6717/\rm{[S\ {\scriptstyle II}]}\lambda6731$ is the emission-line flux ratio of the [S~{\sc ii}] doublet. Coefficients of $a,$ $b,$ and $c$ are the best fit parameters assuming the electron temperature of 10,000 K, which are 0.4315, 2,107.0 and 627.1 respectively. The uncertainty in the density estimate introduced by this temperature assumption is smaller than the typical measurement uncertainty for our sample. 

\begin{figure}
 \begin{center} 
 \includegraphics[width=80mm]{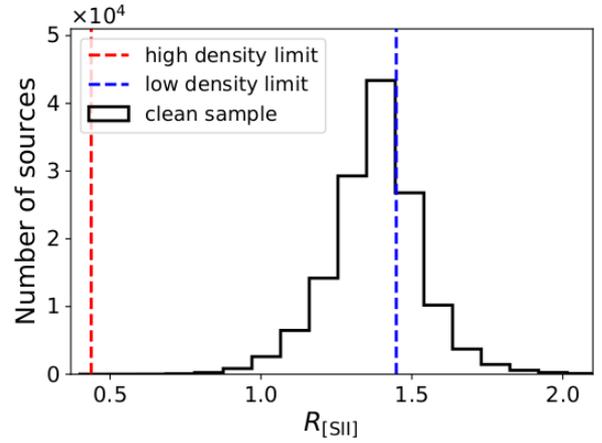}
 \end{center}
 \caption{Histogram of the [S~{\sc ii}] flux ratio ($R_{\rm[S\ {\scriptstyle II}]}$) for the SDSS clean sample. The blue and red dashed lines denote the maximum and minimum flux ratios, respectively (see the main text).}
 \label{sii_hist}
\end{figure}

The expected [S~{\sc ii}] flux ratio is 1.4484 and 0.4375 for electron densities of $1\ \rm{cm^{-3}}$ and $100,000\ \rm{cm^{-3}}$ respectively, which are regarded as the maximum and minimum limits of the [S~{\sc ii}] flux ratio in this work.
\footnote{The gas density of some NLR clouds significantly exceeds 10$^4$ cm$^{-3}$ in some specific situations such as at the innermost part of the NLR and in the high-redshift Universe \citep[e.g.,][]{Fer97, Nag01, Ara12}. Though such high-density clouds are not traced by the [S~{\sc ii}] doublet, we consider that such high-density NLR clouds are not abundant at the host-galaxy scale in the low-redshift Universe.}
In our sample, 41,869 galaxies show larger [S~{\sc ii}] flux ratio than the maximum limit, and 4 galaxies show smaller [S~{\sc ii}] flux ratio than the minimum limit (figure~\ref{sii_hist}). The reason for a relatively large number of galaxies with the [S~{\sc ii}] flux ratio higher than the maximum theoretical value is that the electron density in many emission-line galaxies is too low to be traced by the [S~{\sc ii}] flux ratio. More specifically, 37,308 among 110,041 star-forming galaxies ($\sim$34\%), 3,238 among 18,523 composite galaxies ($\sim$17\%), 874 among 9,571 Seyfert galaxies ($\sim$9\%), and 449 among 2,185 LINERs ($\sim$21\%) show the [S~{\sc ii}] flux ratio exceeding the maximum limit. Thus the typical gas density of ionized clouds in about one third of star-forming galaxies and LINERs is very low, while the fraction of Seyfert galaxies showing such a low density is much lower.
A very minor fraction of the SDSS clean sample (4 among 140,320 emission-line galaxies; $\sim$0.003\%) are characterized by a very high density, but our visual inspections of their spectrum suggest that such a very high gas density is due to the failure of the fit of emission lines by the SDSS pipeline. Such a low failure rate of the fit by the SDSS pipeline is negligibly low in the analysis given in the following sections.

\subsubsection{AGN activity}
Generally it is difficult to assess the AGN luminosity for type-2 AGNs because their central engine is obscured by optically-thick dusty tori. However the AGN activity affects the ionization of NLR gas clouds, which can be traced by some emission-line flux ratios. Some previous works studied how photoionization models reproduce emission-line flux ratios of AGNs on the BPT diagrams \citep[e.g.,][]{Ho93,Gro04}. Such models show that a higher flux ratio of $\rm{[O\ {\scriptstyle III}]\lambda5007/H\beta}$ indicates a higher ionization parameter ($U$), which is the density ratio of the ionizing photon from the AGN to the hydrogen at the irradiated face of NLR gas clouds. In this work, we investigate the electron density and $\rm{[O\ {\scriptstyle III}]}$ velocity dispersion of NLR clouds as a function of the location in the BPT diagram to see how the AGN activity (inferred by the $U$ parameter) affects the properties of NLR gas clouds.

\subsection{The Mapping Nearby Galaxies at the Apache Point Observatory (MaNGA) survey data}
\subsubsection{Sample selection}
We use the MaNGA survey data to investigate spatially resolved physical properties of NLRs. The MaNGA survey is one of the three core programs of SDSS-IV \citep{Bla17} and its goal is to map the detailed composition and kinematic structure of $\sim$10,000 nearby galaxies \citep{Bun15}, obtained with 2$^{\prime\prime}$-diameter fibers. The spectral coverage is 3,600~\AA\ $\lesssim \lambda \lesssim$ 10,000~\AA\ with a spectral resolution of $R\sim2,000$. Each target is observed by a fiber bundle that consists of 19--127 fibers, which corresponds to the diameter of 12--32 arcsec. This field-of-view covers 1.5 $R_{\rm{e}}$ (effective radius) and 2.5 $R_{\rm{e}}$ of targets classified in the Primary and Secondary samples ($\sim$67\% and $\sim$33\% among all MaNGA targets) respectively. In this study, we use both of these two samples. To obtain maps of emission-line fluxes and velocity dispersions, we use the advanced products of the MaNGA Data Analysis Pipeline \citep[DAP;][]{Wes19}. The MaNGA DAP is the survey-led software package that analyzes the data produced by the MaNGA data-reduction pipeline \citep[DRP;][]{Law16} to reproduce physical properties derived from the MaNGA spectroscopy.

We select the sample from the DAPALL catalog made by the MaNGA DAP, according to the following procedure. The objects with DAPDONE = False, in this case DAP was not successful, are excluded. After removing the duplicated objects, we obtain the DAP outputs of 4,609 galaxies. Note that the DAP provides two kinds of maps with a different spatial sampling. One of the two is VOR10, in which spaxels are binned to achieve S/N$\sim$10 based on the Voronoi binning algorithm \citep{Cap03}. The other one is hybrid binning scheme (HYB), in which the emission-line measurement is performed on the original spaxel sampling in the spatial dimension. In this study, we adopt HYB since it provides better spatial sampling for the analysis. 

\subsubsection{The central and off-central regions of MaNGA galaxies}
In case that galaxies are in the dynamical equilibrium, the velocity dispersion of emission lines from ionized gas clouds is expected to be proportional to the gravitational potential of galaxies. However, if the outflow (either of starburst-driven and AGN-driven) disturbs the ISM of a galaxy, the velocity dispersion is consequently enhanced. In this study, we compare the velocity dispersion of the [O~{\sc iii}] line for Seyfert galaxies and star-forming galaxies with a similar stellar mass. To classify the MaNGA objects, we match the MaNGA sample (4,609 objects) and the SDSS clean sample (140,320 objects). Here, we focus specifically on star-forming galaxies and Seyfert galaxies to see possible effects of the AGN activity onto the ISM. As a result, 90 and 801 MaNGA objects are classified as Seyfert and star-forming galaxies respectively. As for the stellar mass, we adopt the median of the stellar-mass probability distribution given in the MPA-JHU catalog.

\begin{figure}
 \begin{center}
 \includegraphics[width=78mm]{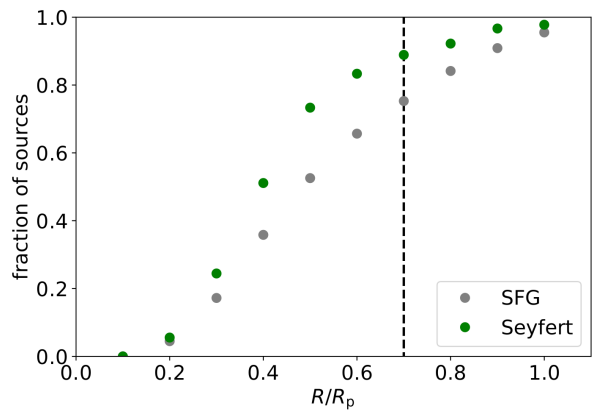}
 \end{center}
 \caption{The fractions of sources whose angular radius from the center on the minor axis ($R$) is larger than 1.5 arcsec, as a function of $R/R_{\rm{p}}$. Gray and green dots denote 801 star-forming galaxies and 90 Seyfert galaxies, respectively. The vertical dashed line corresponds to $R/R_{\rm p} = 0.7$ (see the main text).
 }
 \label{seeing}
\end{figure}

Here we define the central and off-central regions of MaNGA galaxies, to investigate the possible effect of outflows at inner and outer parts of galaxies in our sample. The projected galaxy size is characterized by the two-dimensional Petrosian radius $R_{\rm p}$ \citep{Bla17, Wes19} in the DAP output. We define the off-central region as described below, so that the nuclear emission does not significantly contribute to the off-central region due to the seeing effect in most objects. Figure~\ref{seeing} shows the fractions of sources whose angular radius from the center on the minor axis ($R$) is larger than $\sim$1.5 arcsec, as a function of $R/R_{\rm p}$, for 801 star-forming galaxies and 90 Seyfert galaxies. Since the typical size of the point-spread function (PSF) of the MaNGA data is $\sim$2.5 arcsec in diameter, the boundary between the central and off-central regions should be located at 1.5 arcsec from the nucleus at least. Figure~\ref{seeing} shows that the radius from the nucleus on the minor axis at $R/R_{\rm{p}}=0.7$ is larger than 1.5 arcsec in most cases ($\sim$89\% of Seyfert galaxies and $\sim$75\% of star-forming galaxies). Therefore, we define the region between the ellipses with $R/R_{\rm{p}}=$ 0.7 and 1.0 as the off-central region. Then we define the region within the ellipse with $R/R_{\rm{p}}=$ 0.7 as the central region. By this definition, the number of spaxels in central and off-central regions becomes almost the same; i.e., roughly the half of the area within the ellipse with $R/R_{\rm{p}}=$ 1.0 belongs to the central region ($R/R_{\rm{p}} < 0.7$) and the remaining half belongs to the off-central region ($0.7 < R/R_{\rm{p}} < 1.0$), in each object.

\begin{figure*}
 \begin{center}
  \includegraphics[width=175mm]{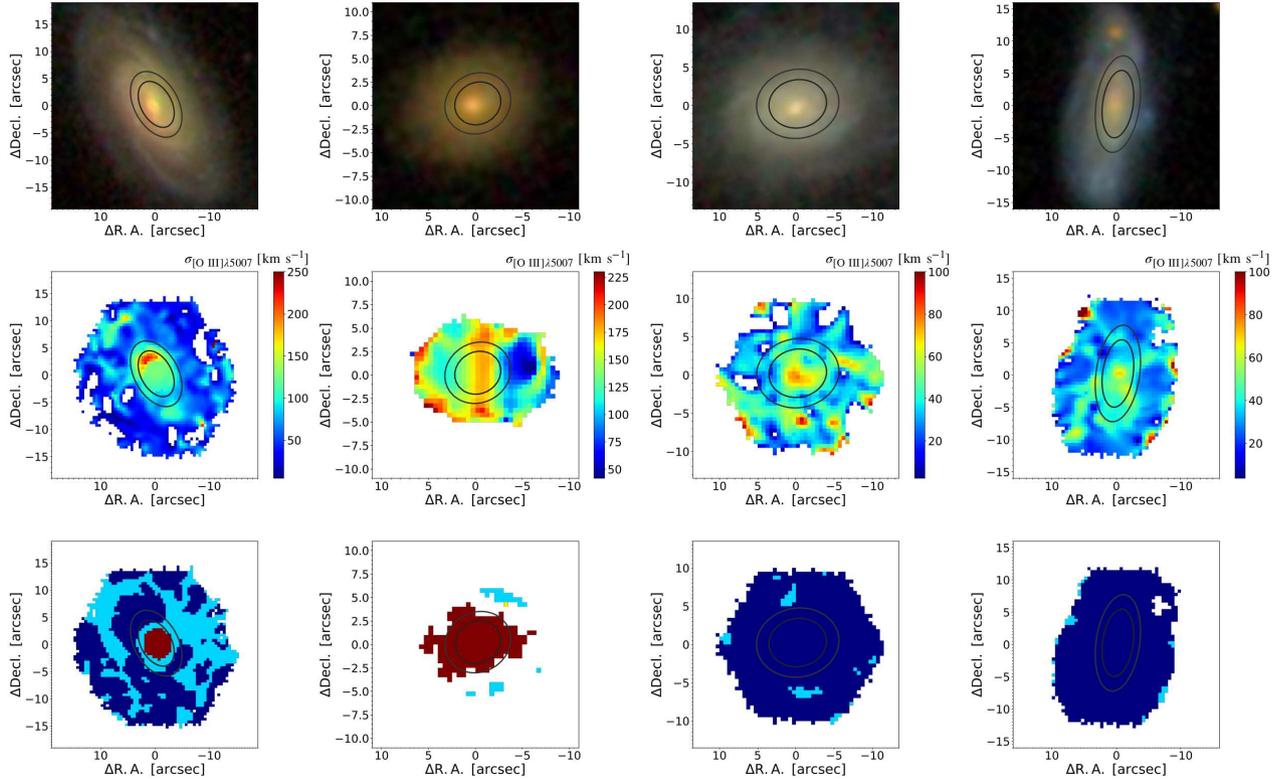}
 \end{center}
 \caption{Examples of SDSS cutout $gri$ three-color images (upper panels), $\rm{[O\ {\scriptstyle III}]}$ velocity dispersion maps obtained by the MaNGA survey (middle panels), and BPT maps with NLR, H~{\sc ii} region, LINER, and composite region, colored by brown, blue, yellow, and green (lower panels) for two Seyfert galaxies (left two columns) and two star-forming galaxies (right two columns). The ellipses with $R/R_{\rm{p}}=$ 0.7 and 1.0 are shown in each panel.}
 \label{example}
\end{figure*}

Figure \ref{example} shows the examples of three-color SDSS images with $g$-, $r$-, and $i$-bands, and $\rm{[O\ {\scriptstyle III}]}$ velocity dispersion maps for Seyfert and star-forming galaxies, with the ellipses defining the central and off-central regions. For examining a possible contribution of the outflow to the gas dynamics, we measure the velocity dispersion of the central and off-central regions by calculating the 75th percentile of the [O~{\sc iii}] velocity dispersion of all spaxels at $R/R_{\rm{p}}<0.7$ and $0.7<R/R_{\rm{p}}<1.0$, not the median nor average. This is because the AGN-driven and starburst-driven outflows are sometimes collimated and thus they affect the dynamics of gas only in the (bi-)polar direction with a small filling factor \citep[e.g.,][]{Hec90, Fal98, Sun18}. Consequently, the median and average of the velocity dispersion may not be good indicators to examine the possible contribution of the outflow. In some cases, the number of available spaxels in the central or off-central regions is too small to characterize the velocity dispersion, due to the too narrow size of the region, a large number of low-S/N ($< 3$) or masked-out spaxels, or the combination of them. Therefore we remove the objects with the number of available spaxels less than 10 for both of the central region and the off-central region.

Using $\rm{[O\ {\scriptstyle III}]}$ velocity dispersion for the central region (in 794 star-forming galaxies and 88 Seyfert galaxies) and the off-central region (in 795 star-forming galaxies and 89 Seyfert galaxies) derived by the above method, we will compare the relation between stellar mass and velocity dispersion of Seyfert galaxies and star-forming galaxies (section~3.3).

Figure~\ref{example} also shows the maps of the BPT classification of emission-line regions (NLR, H~{\sc ii} region, and composite region) for each spaxel. In most star-forming galaxies, the spaxels in both central and off-central regions are classified as the H~{\sc ii} region. Though the spaxels both in central and off-central regions of many Seyfert galaxies are classified as the NLR (as the second object in figure~\ref{example}), not all spaxels of some Seyfert galaxies are classified as the NLR, especially in off-central regions (as the first object in figure~\ref{example}). This should be kept in mind during the analyses given in the following sections.

\section{Results}
\subsection{Statistical properties of gas in emission-line galaxies}

\begin{figure}
 \begin{center}
 \includegraphics[width=75mm]{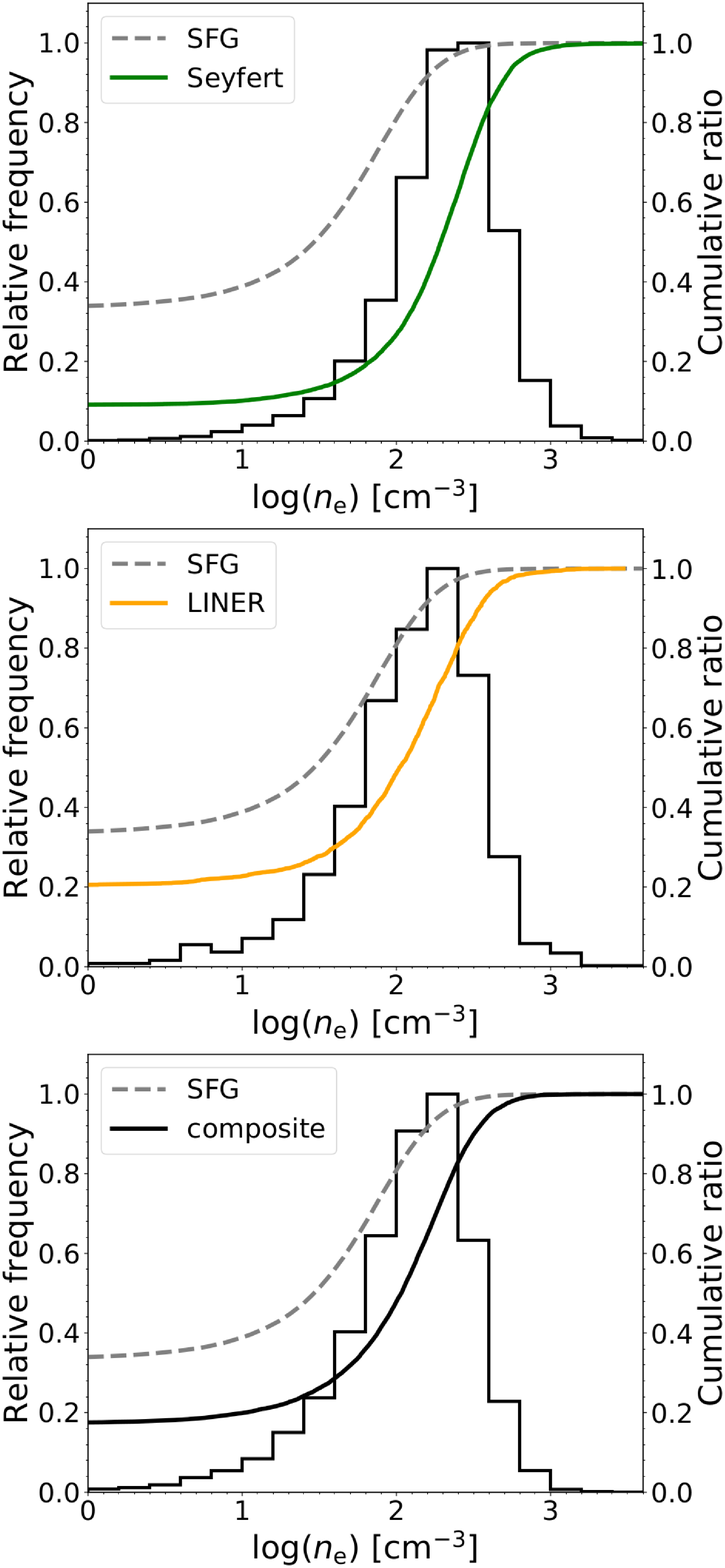}
 \end{center}
 \caption{Histograms and cumulative distributions of the electron density for Seyfert galaxies, LINERs, and composite galaxies are shown from top to bottom panels, respectively. The gray-dashed line in all panels shows the cumulative distribution for the star-forming galaxies. The cumulative distributions start from non-zero value, because of the presence of objects whose [S~{\sc ii}] flux ratio is higher than the low-density limit.}
  \label{ne_hist}
\end{figure}

\begin{figure}
 \begin{center}
 \includegraphics[width=75mm]{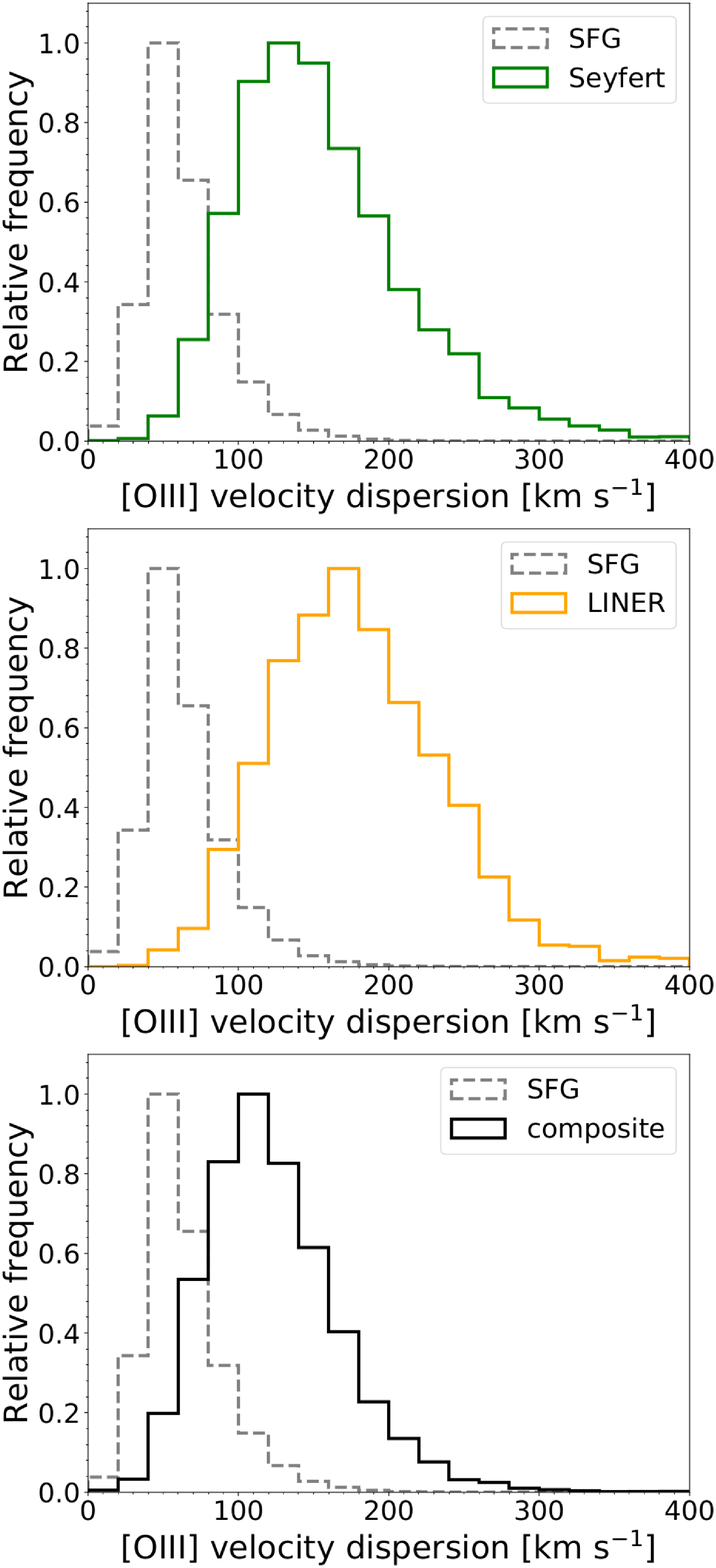}
 \end{center}
 \caption{Histograms of the $\rm{[O\ {\scriptstyle III}]}$ velocity dispersion for Seyfert galaxies, LINERs, and composite galaxies are shown from top to bottom panels, respectively.  The histogram drawn by gray-dashed line shows the distribution for star-forming galaxies.}\vspace{7mm}
 \label{sigma_hist}
\end{figure}

Table~\ref{table_ne} shows the median of the electron density and the $\rm{[O\ {\scriptstyle III}]}$ velocity dispersion measured by the SDSS fiber spectroscopy, for each class of emission-line galaxies. Since many objects show the [S~{\sc ii}] flux ratio deviated from the range between the low-density and high-density limits as shown in section 2.1.2 (see also figure~\ref{sii_hist}), table~\ref{table_ne} shows the median of the electron density in cases of both including and excluding those objects. In both cases, LINERs and composite galaxies show similar density. Compared with these two classes (LINERs and composite galaxies), Seyfert galaxies and star-forming galaxies show systematically higher and lower gas density, respectively. For clarifying these trends, the histograms of the electron density for each class of emission-line galaxies are shown in figure~\ref{ne_hist}. Since those histograms cannot show objects out of the range between the low- and high-density limits, the cumulative distribution function of the gas density is also shown in this figure. The modes of the distribution of the gas density (cm$^{-3}$) in the logarithmic scale are 2.5 and 1.9 for Seyfert galaxies and star-forming galaxies, i.e., the gas density in Seyfert galaxies is $\sim$0.6 dex higher than that in star-forming galaxies. For examining the statistical significance of the difference, we apply the Kolmogorov-Smirnov (KS) statistical test for the frequency distributions of the gas density of Seyfert galaxies and star-forming galaxies. The KS test shows that the probability of these two distributions taken from the same parent distribution is lower than 10$^{-10}$, suggesting that the difference in the density distribution between Seyfert galaxies and star-forming galaxies is statistically significant.

\begin{table}[htb]
   \caption{The median values of the electron density and $\rm{[O\ {\scriptstyle III}]}$ 
   velocity dispersion for each class of emission-line galaxies}
   \begin{tabular}{lcccc} \hline
     classification & 
     Number & 
     $n_{\rm{e}}$\_A${}^{*}$ &
     $n_{\rm{e}}$\_B${}^{\dag}$ & 
     $\sigma_{\rm [OIII]5007}$ \\ 
     &  
     & 
     [cm$^{-3}$] & 
     [cm$^{-3}$] & 
     [km s$^{-1}$] \\ \hline\hline
     Seyfert & 9,571 & 194 & 213 & 147 \\
     LINER & 2,185 & 104 & 146 & 173 \\ 
     SF & 110,041 & 29 & 61 & 58 \\
     composite & 18,523 & 108 & 136 & 118 \\ \hline
   \end{tabular}
  \label{table_ne}
  \begin{tabnote}
$^*$ Median value measured including objects with the density above the high-density limit and those below the low-density limit.\\
$^\dag$ Median value measured excluding objects with the density above the high-density limit and those below the low-density limit.
\end{tabnote}
\end{table}

Based on table~\ref{table_ne}, it is suggested that the [O~{\sc iii}] velocity dispersion of Seyfert galaxies and LINERs is much larger than that of star-forming galaxies, and the [O~{\sc iii}] velocity dispersion of composite galaxies is intermediate among them. Figure~\ref{sigma_hist} shows the frequency distribution of the [O~{\sc iii}] velocity dispersion for each class of emission-line galaxies, displaying the consistent trend. The modes of the distribution of the [O~{\sc iii}] velocity dispersion of Seyfert galaxies and star-forming galaxies are 130 km s$^{-1}$ and 50 km s$^{-1}$ respectively, i.e., the [O~{\sc iii}] velocity dispersion in Seyfert galaxies is $\sim$0.4 dex larger than that in star-forming galaxies. The KS test shows that the probability of these two distributions taken from the same parent distribution is lower than 10$^{-10}$, so the difference in the distribution of the [O~{\sc iii}] velocity dispersion between Seyfert galaxies and star-forming galaxies is also statistically significant.

\subsection{The properties of NLRs as a function of a location in the BPT diagram}
\begin{figure}
 \begin{center}
 \includegraphics[width=80mm]{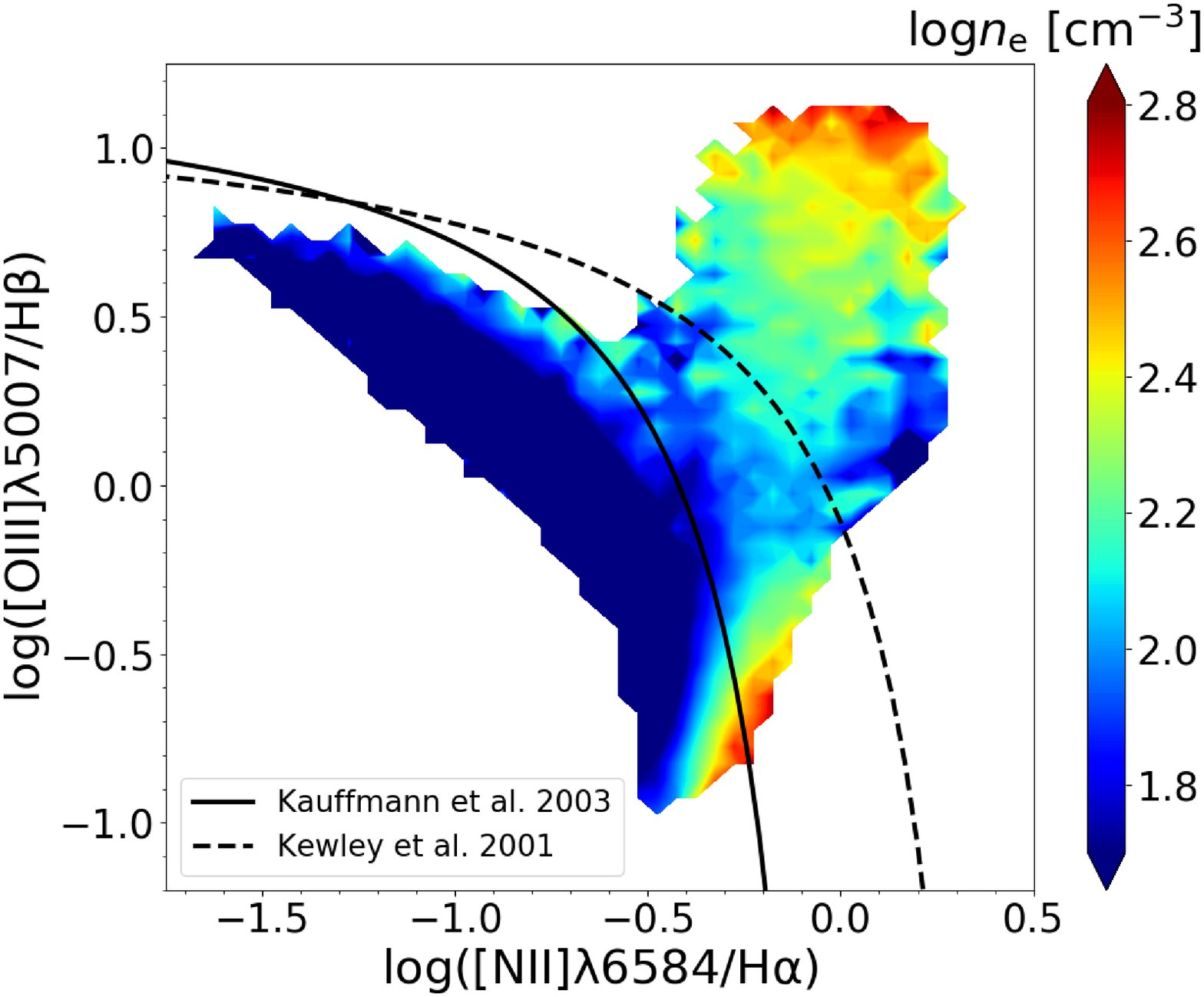} 
 \end{center}
 \caption{Contour of the electron density on the BPT diagram for all galaxies in the SDSS clean sample. The median electron density for each 0.05 dex $\times$ 0.05 dex bin is shown with a color, whose scale is shown in the right side of the panel. Bins with $<5$ galaxies are not shown. The dashed line and solid line on the top panel show the limitation line showing the region star-forming galaxies distribute by \citet{Kew01} and the empirical line by \citet{Kau03b} separating star-forming galaxies and composite galaxies respectively.}
 \label{bpt_ne}
\end{figure}

In figure \ref{bpt_ne}, we show the electron density of ionized gas clouds as a function of the location in the BPT diagram. This figure is drawn by calculating the median value for each grid with a width of 0.05 dex on the BPT diagram. Here, the objects with the [S~{\sc ii}] flux ratio deviated from the range between the low-density and high-density limits are also included for drawing the contour of the electron density. In the region where star-forming galaxies distribute, the electron density is low, typically $n_{\rm e} < 10^2$ cm$^{-3}$. On the other hand, in the region where AGNs (Seyfets galaxies and LINERs) and composites distribute, the value of the electron density shows $n_{\rm e} \sim 10^{2-3}$ cm$^{-3}$. Interestingly, we found that the electron density is higher at 
the region with the higher ratio $\rm{[O\ {\scriptstyle III}]}\lambda5007/\rm{H}\beta$ (i.e. the upper-right side in the BPT diagram), reaching up to $\sim$10$^{2.7}$ cm$^{-3}$ at $\log(\rm{[O\ {\scriptstyle III}]}\lambda5007/\rm{H}\beta) \sim 1$. In addition, objects characterized by a systematically high electron density are seen also at the lower edge of the distribution of composite galaxies, showing $n_{\rm e} \gtrsim 10^{2.5}$ cm$^{-3}$.

\begin{figure}
 \begin{center}
 \includegraphics[width=80mm]{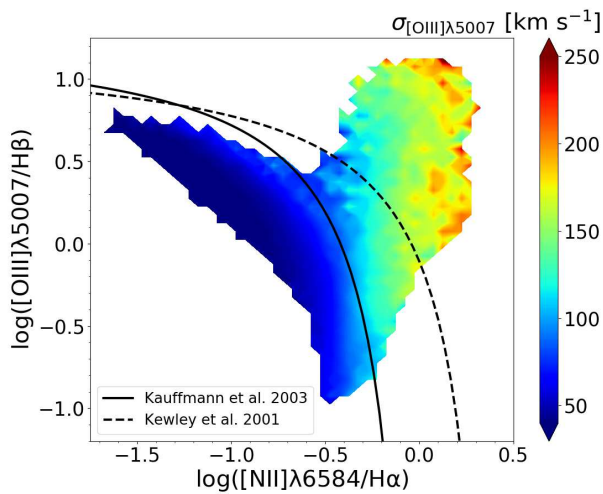}
 \end{center}
 \caption{Same as figure~\ref{bpt_ne} but for the velocity dispersion of the $\rm{[O\ {\scriptstyle III}]}\lambda5007$ emission line.}
 \label{bpt_sigma}
\end{figure}

Next, we investigate the $\rm{[O\ {\scriptstyle III}]}$ velocity dispersion as a function of the location in the BPT diagram (figure~\ref{bpt_sigma}). Below the criterion of \citet{Kau03a}, the velocity dispersion is low, typically less than $\sigma_{\rm{[O\ {\scriptstyle III}]}}=100\ \rm{km\ s^{-1}}$. The velocity dispersion is higher at the upper right direction in the BPT diagram, reaching up to $\sim$250 km s$^{-1}$ at $\log(\rm{[O\ {\scriptstyle III}]}\lambda5007/\rm{H}\beta) \sim 1$. At $\log$([N~{\sc ii}]/H$\alpha$) $>$ 0.0 and $\log$([O~{\sc iii}]/H$\beta$) = 0.0 -- 0.5 where LINERs are distributed, the velocity dispersion is moderately high ($\sim$200 km s$^{-1}$) though the electron density is low (figure~\ref{bpt_ne}). The high velocity dispersion of LINERs is naturally expected by the fact that LINERs are preferentially seen in massive elliptical galaxies \citep[e.g.,][]{Hec80, Ho03} and sometimes associated with fast shocks \citep[e.g.,][]{Dop97, Ric10, Ric11}. At the lower edge of the distribution of composite galaxies, the velocity dispersion is not specifically high (i.e., $\sim$100 km s$^{-1}$), which is contrary to the distribution of the electron density shown in figure~\ref{bpt_ne}.

\begin{figure*}
 \begin{center}
 \includegraphics[width=172mm]{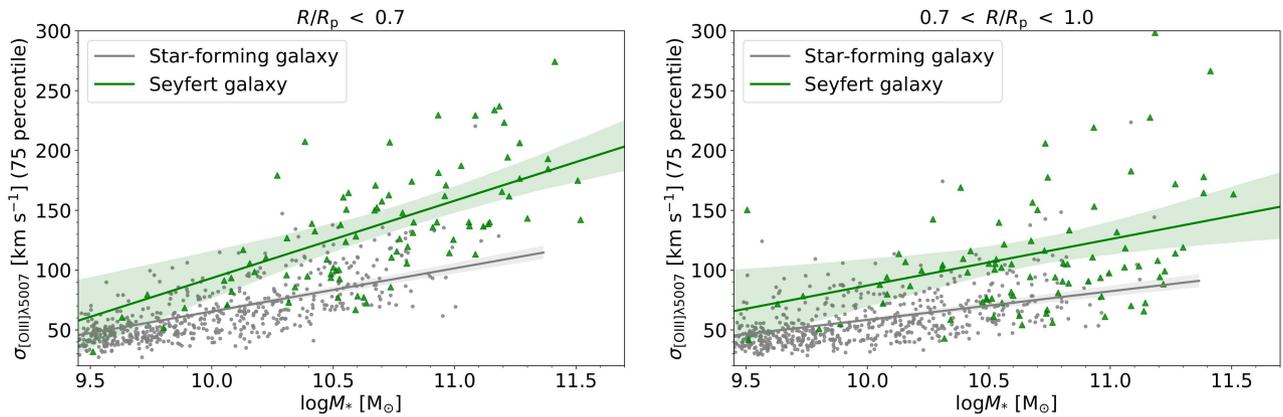}
 \end{center}
 \caption{Velocity dispersion of $\rm{[O\ {\scriptstyle III}]}\lambda5007$ vs. stellar mass for 90 Seyfert galaxies and 801 star-forming galaxies from MaNGA sample. In the left panel, the $\rm{[O\ {\scriptstyle III}]}$ velocity dispersion is derived by measuring 75th percentile of all spxels in the central region ($R/R_{\rm{p}}<0.7$). On the other hand, in the right panel, the $\rm{[O\ {\scriptstyle III}]}$ velocity dispersion is derived by measuring 75th percentile of all spxels in the offset region ($0.7<R/R_{\rm{p}}<1.0$). The green line and grey one show the regression line for Seyfert galaxies and star-forming galaxies respectively. The green shade and grey shade show the 95\% confidence interval estimated by using the bootstrap method for Seyfert galaxies and star-forming galaxies respectively.}
 \label{sigma_mass}
\end{figure*}

\subsection{$\rm{[O\ {\scriptstyle III}]}$ velocity dispersion vs. stellar mass in the central and off-central regions}
In section~3.2, it is clearly shown that the velocity dispersion is larger in Seyfert galaxies than in star-forming galaxies (figure~\ref{bpt_sigma}). However it is unclear whether this is caused by the AGN outflow, because the velocity dispersion is largely affected by the stellar mass of galaxies and thus the mass-matched comparison is needed to study the effect of the AGN outflow. Even if the larger velocity dispersion of Seyfert galaxies is attributed by the AGN outflow, it is unclear where the AGN outflow makes the velocity dispersion larger in their host galaxies because the SDSS fiber spectroscopic data lack the spatial information. Therefore, we study the velocity dispersion of ionized gas clouds by using the spatially-resolved spectroscopic data of MaNGA for 90 Seyfert galaxies and 801 star-forming galaxies (section~2.2.2), and compare them in a mass-matched way. 

Figure~\ref{sigma_mass} shows the relation between the $\rm{[O\ {\scriptstyle III}]}$ velocity dispersion and the stellar mass for the Seyfert galaxies and the star-forming galaxies on the central region (the left panel; $R/R_{\rm{p}}<0.7$) and the off-central region (the right panel; $0.7<R/R_{\rm{p}}<1.0$), respectively. For both of Seyfert galaxies and star-forming galaxies, the positive correlation between the $\rm{[O\ {\scriptstyle III}]}$ velocity dispersion and the stellar mass is seen for both the central and off-central regions, as expected. Interestingly, at a given stellar mass, Seyfert galaxies show a larger velocity dispersion than star-forming galaxies systematically, for both the central and off-central regions. This is clearly shown by the regression lines in figure~\ref{sigma_mass}. This result is consistent to a previous work by \citet{Woo16}, who reported an excess of the gas velocity dispersion with respect to the stellar velocity dispersion for SDSS type-2 Seyfert galaxies. In the central region ($R/R_{\rm{p}}<0.7$), the slope of the regression line is steeper for Seyfert galaxies than that for star-forming galaxies, suggesting that the excess in the velocity dispersion of Seyfert galaxies is more significant in objects with a larger stellar mass. Here we note that the results do not depend significantly on the threshold to separate the central and off-central regions. The obtained results are qualitatively the same if we adopt $R/R_{\rm{p}}=$ 0.3, 0.5, or 0.8, instead of $R/R_{\rm{p}}=0.7$ as the boundary between the central and off-central regions.

The [O~{\sc iii}] velocity dispersion of one Seyfert galaxy (Mrk~622) is too large to be displayed within the displayed range of figure~\ref{sigma_mass}. The stellar mass of Mrk~622 is $\log (M_*/M_\odot) = 10.3$, and its velocity dispersions at central and off-central regions are $\sigma_{\rm{[O\ {\scriptstyle III}]}}$ = 455 km s$^{-1}$ and 411 km s$^{-1}$, respectively. Figure \ref{mrk} shows the MaNGA $\rm{[O\ {\scriptstyle III}]}$ velocity dispersion map of Mrk~622. The velocity dispersion is very large (400--500 km s$^{-1}$) at almost all spaxels in the central and off-central regions, i.e., not only at the nucleus of Mrk~622. {\citet{Ben19} found that Mrk~622 has double or triple AGNs, which suggests that the galaxy has experienced a merger. Such merging events may enhance the velocity dispersion of gas clouds in the entire region of Mrk~622. Thus this object is very interesting to study the relation between the galaxy merger and AGN activity. We do not discuss this object further because this is beyond the scope of this paper.

\begin{figure}
 \begin{center}
 \includegraphics[width=85mm]{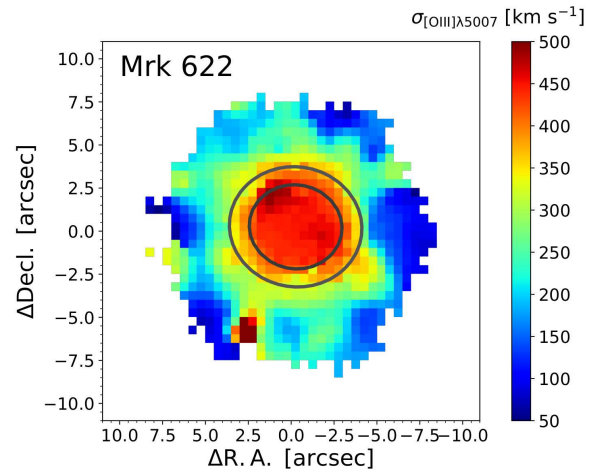}
 \end{center}
 \caption{The $\rm{[O\ {\scriptstyle III}]}$ velocity dispersion map of the Seyfert galaxy Mrk 622. The ellipses with $R/R_{\rm{p}}=$ 0.7 and 1.0 are showed on this map.}
 \label{mrk}
\end{figure}

\section{Discussion}
\subsection{Where do gas clouds in narrow-line regions come from?}
In this study, using the large spectroscopic data set of the SDSS, we showed that the electron density and [O~{\sc iii}] velocity dispersion of gas clouds in NLRs of Seyfert galaxies are systematically higher than those in H~{\sc ii} regions of star-forming galaxies. Among Seyfert galaxies, objects with a higher [O~{\sc iii}]/H$\beta$ flux ratio (inferring a larger ionization parameter) show higher electron density and [O~{\sc iii}] velocity dispersion. Here, we note that the trend of the velocity dispersion may be due to the bias of the stellar mass, because the stellar mass of AGNs is systematically higher than that of star-forming galaxies in our sample. Specifically, the averages and standard deviations of the stellar mass of the selected star-forming galaxies, composite galaxies, Seyfert galaxies, and LINERs are $\log{(M_{*}/M_{\odot})} = 9.84\pm0.58$, $\log{(M_{*}/M_{\odot})} = 10.59\pm0.37$, $\log{(M_{*}/M_{\odot})} = 10.73\pm0.35$ and $\log{(M_{*}/M_{\odot})} = 10.90\pm0.36$, respectively. To examine this concern, we separate the SDSS clean sample to stellar mass bins with $\Delta \log{(M_\ast/M_\odot)} = 0.5$, and then make a contour of the [O~{\sc iii}] velocity dispersion on the BPT diagram for each bin. As shown in figure~\ref{bpt_sigma_bin}, the [O~{\sc iii}] velocity dispersion of AGNs is systematically larger than that of star-forming galaxies by a factor of $>2$ even when objects with a similar stellar mass are 72.1 km~s$^{-1}$ and 117.6 km~s$^{-1}$ for the range of 10.0 $<$ log ($M_\ast/M_\odot$) $<$ 10.5 and 98.2 km~s$^{-1}$ and 148.8 km~s$^{-1}$ for 10.5 $<$ log ($M_\ast/M_\odot$) $<$ 11.0, respectively. Note that the [O~{\sc iii}] velocity dispersion of Seyfert galaxies with 10.0 $<$ log ($M_\ast/M_\odot$) $<$ 10.5 is larger than that of star-forming galaxies with 10.5 $<$ log ($M_\ast/M_\odot$) $<$ 11.0 (see also figure 11), suggesting that the higher velocity dispersion of Seyfert galaxies compared to star-forming galaxies cannot be explained only by the effect of the stellar mass.

\begin{figure}
 \begin{center}
 \includegraphics[width=85mm]{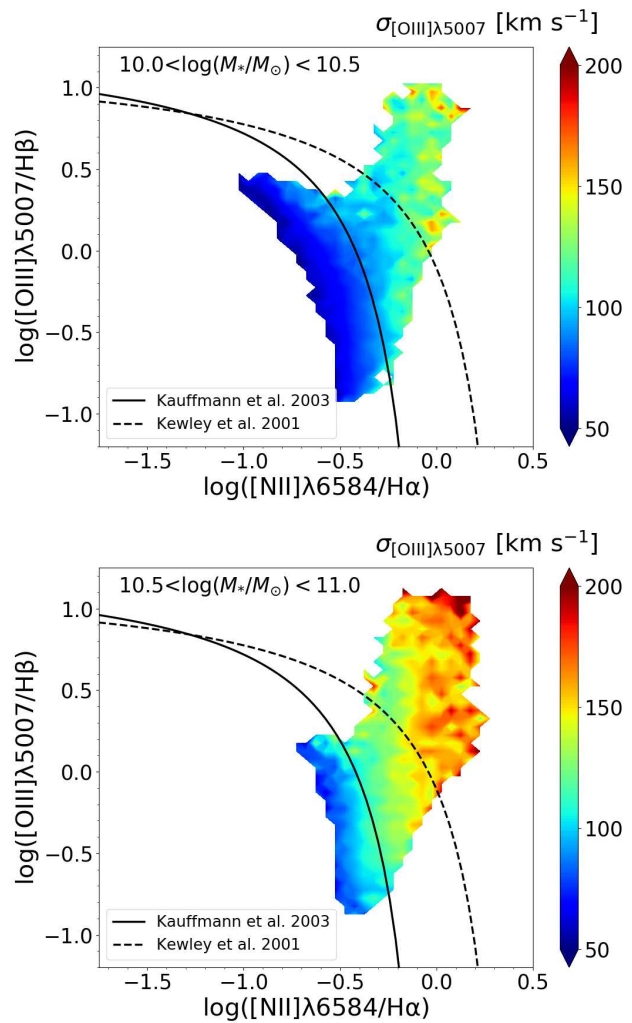}
 \end{center}
 \caption{Same as figure~\ref{bpt_sigma}, but only for galaxies with $10.0<\log{(M_{*}/M_{\odot})}<10.5$ (upper panel) and $10.5<\log{(M_{*}/M_{\odot})}<11.0$ (lower panel), respectively.}
 \label{bpt_sigma_bin}
\end{figure}

The higher gas density and velocity dispersion of NLRs in Seyfert galaxies than those of H~{\sc ii} regions in star-forming galaxies cannot be easily explained by a picture that the origin of the NLR in Seyfert galaxies is gas clouds originally located there and ionized by the nuclear ionizing radiation. Instead, the observed properties of the NLR are more naturally understood by a scenario that the NLR gas clouds are originated at much closer part of their nucleus, plausibly transferred by the AGN outflow. \citet{Wad18} theoretically showed that the spectral properties and the morphology of the NLR in the Circinus galaxy (a nearby archetypal type-2 Seyfert galaxy) can be explained by their radiation-driven fountain model, where the NLR gas clouds are transferred from the nucleus through the radiation-driven AGN outflow. In this model, the gas density and ionization parameter of the simulated NLR are typically $n_{\rm e} \sim$ 300--1,500 cm$^{-3}$ and $U \sim$0.01. The $U$ parameter predicted by \citet{Wad18} corresponds to $\log({\rm [O\ {\scriptstyle III}]\lambda5007/\rm H\beta})\sim1$ \citep[e.g.][]{Gro04}, and therefore the simulated NLR properties are similar to the observed NLR properties of Seyfert galaxies specifically at the upper-right side in the BPT diagram (section~3.2; see also figure~\ref{bpt_ne}). 
Though this similarity supports the idea that the NLR clouds are originated in the AGN-driven outflow, the simulated NLR by the radiation-driven fountain model of \citet{Wad18} is much more compact ($\sim$10 pc from the nucleus) than the NLRs in our SDSS sample. Further theoretical simulations for NLRs with a larger spatial scale are required to understand the origin of NLR gas clouds in more detail.

\begin{figure*}
 \begin{center}
 \includegraphics[width=173mm]{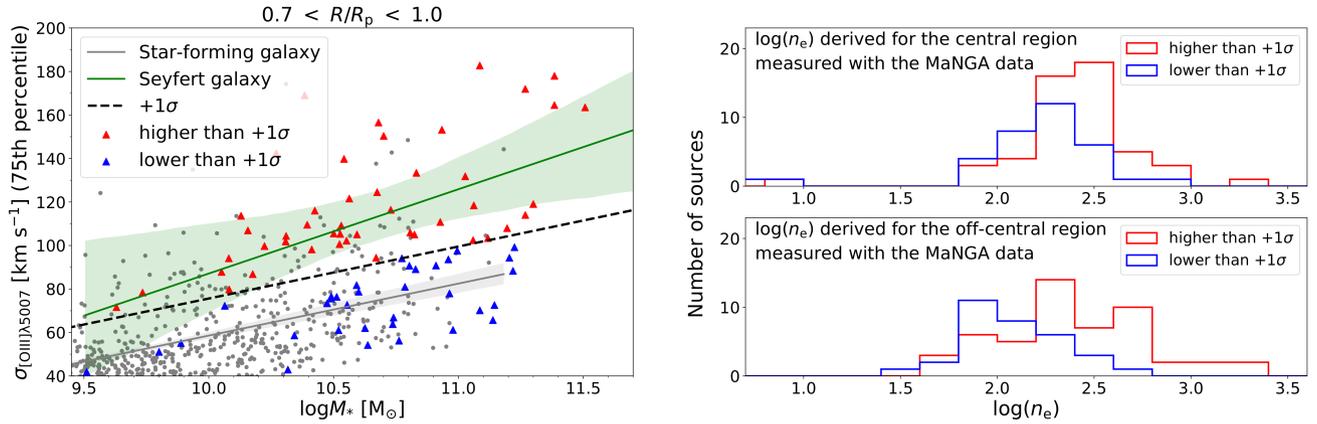}
 \end{center}
 \caption{(Left) Same as the right panel in figure \ref{sigma_mass} but 1 sigma standard deviation above the regression line for star-forming galaxies is shown with a dashed line. (Right) Histograms of the electron density in Seyfert galaxies, derived for the central region ($0.0 < R/R_{\rm{p}} < 0.7$; upper panel) and for the off-central region ($0.7 < R/R_{\rm{p}} < 1.0$; lower panel) measured with the MaNGA data, measured at the 75 percentile. The red and blue lines denote the histograms for Seyfert galaxies whose velocity dispersion is larger and smaller than the +1$\sigma$ relation (the dashed line in the left panel), respectively.
}
\label{sigma_mass_ne}
\end{figure*}

In section~3.3, we showed that the $\rm{[O\ {\scriptstyle III}]}$ velocity dispersion of Seyfert galaxies is systematically larger than that of star-forming galaxies at a given stellar mass. If this is due to the AGN-driven outflow, NLRs with the enhanced velocity dispersion should be characterized by a higher density than NLRs whose velocity dispersion is not large. To test this idea, we define ``Seyfert galaxies with the enhanced velocity dispersion'' as Seyfert galaxies whose [O~{\sc iii}] velocity dispersion at the off-central region is larger than +1$\sigma$ from the typical velocity dispersion of star-forming galaxies (i.e., Seyfert galaxies above the dashed line in the left panel of figure~\ref{sigma_mass_ne}). Here we focus on the off-central region because the enhancement of the velocity dispersion of Seyfert galaxies in the off-central region is expected to be linked to the AGN outflow in the entire host-galaxy scale. According to this definition, the samples of Seyfert galaxies with and without the enhanced velocity dispersion consist of 53 and 34 objects, respectively. The right panel of figure~\ref{sigma_mass_ne} shows the histograms of the electron density for both of these two samples. Here we show the two cases corresponding to different measurements of the electron density; measurements for the MaNGA central region ($R/R_{\rm p} < 0.7$) and off-central region ($0.7 < R/R_{\rm p} < 1.0$) at the 75th percentile of the electron density, respectively. Both cases show that the electron density of the sample with a higher velocity dispersion is systematically higher than that of the other sample. The medians of the electron density distribution in the logarithmic scale ($\rm{cm^{-3}}$) for the samples with and without the enhanced velocity dispersion are 2.42 and 2.25 in the case of the measurement for the MaNGA central region and 2.32 and 2.02 in the case of the measurement for the MaNGA off-central region, respectively. The p-values from the KS test for the two cases of the electron density measurement are 0.0039 and 0.0002, for the MaNGA central and off-central measurement respectively. This suggests that the difference in the electron density is statistically significant in the case of the MaNGA off-central measurement, though the difference is statistically marginal in the case of the MaNGA central measurement. This result is consistent to a picture that the AGN outflow transfers high-density gas clouds from the nucleus to the host-galaxy scale that results in a high electron density and the velocity dispersion at the off-central region of Seyfert galaxies. But it is also suggested that such AGN outflow does not work for all Seyfert galaxies, since some Seyfert galaxies show comparable velocity dispersion to star-forming galaxies with a similar stellar mass, in which cases the electron density of the NLR is not significantly high. The inferred relationship between the gas density and velocity dispersion of NLR clouds seems contrary to the result reported in \citet{Zha13} (see section 1). This discrepancy is probably because the effect of the stellar mass on the velocity dispersion, that is taken into account during our analysis but not in \citet{Zha13}.

\subsection{High density gas in low-excitation composite galaxies}
As shown in figure \ref{bpt_ne}, composite galaxies with a low $\rm{[O\ {\scriptstyle III}]\lambda5007/H\beta}$ flux ratio show a significantly higher electron density, which is comparable to the electron density of Seyfert galaxies with a high $\rm{[O\ {\scriptstyle III}]\lambda5007/H\beta}$ flux ratio. Because the low $\rm{[O\ {\scriptstyle III}]\lambda5007/H\beta}$ flux ratio suggests a small ionization parameter, gas clouds in those composite galaxies are not exposed by the strong AGN power-law ionizing radiation. Therefore it may be difficult to explain the density enhancement in those composite galaxies by the powerful AGN outflow. 

One idea to explain this density enhancement is the gas compression by the radio jet \citep[e.g.,][]{Wag12, Gai12, Cre15}, which is seen more frequently in AGNs with a low activity characterized by a low Eddington ratio \citep[e.g.,][]{Sik07}. To examine this possibility, we investigate the contribution of radio jets by using the data set of the Faint Images of the Radio Sky at Twenty cm survey \citep[FIRST;][]{Bec95} which covers almost all ($\sim$95\%) of the SDSS sky coverage. For simplicity, we here define the radio-loud AGN as an object whose rest-frame 1.4 GHz radio luminosity is higher than $10^{23}$ W Hz$^{-1}$ \citep{Bes05}. Figure \ref{bpt_radio} shows the fraction of radio-loud objects as a function of a location in the BPT diagram. First, it is clearly shown that the radio-loud fraction of star-forming galaxies is very low, as expected. A high radio-loud fraction is seen in Seyfert galaxies with a large $\rm{[O\ {\scriptstyle III}]\lambda5007/H\beta}$ flux ratio (located at the upper-right edge in the BPT diagram) and LINERs with a small $\rm{[O\ {\scriptstyle III}]\lambda5007/H\beta}$ flux ratio (at the lower-right edge in the BPT diagram). These two populations of radio-loud AGNs are plausibly corresponding to the dichotomy of high- and low-excitation radio galaxies \citep[e.g.,][]{Bes12,Jan12,Pra16}, given the clear difference in the $\rm{[O\ {\scriptstyle III}]\lambda5007/H\beta}$ flux ratio. Thus the observed high gas density of Seyfert galaxies with a large [O~{\sc iii}]/H$\beta$ ratio (see figure 7) may be partly affected also by the gas compression due to the radio jet, not only by the transfer of high-density gas clouds from the nucleus (see section 4.1). On the other hand, the radio-loud fraction of composite galaxies with a low $\rm{[O\ {\scriptstyle III}]\lambda5007/H\beta}$ flux ratio is quite low, suggesting that the AGN radio jet is not the main contributor to the enhanced electron density of low-excitation composite galaxies. For understanding the physical origin of the density enhancement in low-excitation composite galaxies, more thorough analyses based on multi emission-line diagnostics including much fainter lines than the standard BPT lines will be required. 

\begin{figure}
 \begin{center}
 \includegraphics[width=80mm]{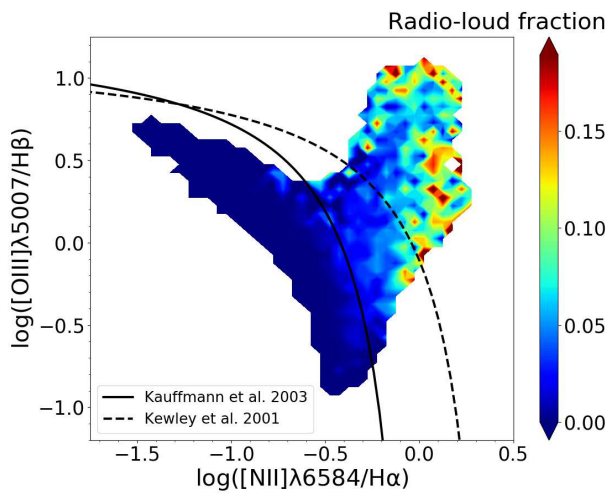}
 \end{center}
 \caption{Same as figure~7 but for the radio-loud fraction.}
 \label{bpt_radio}
\end{figure}

\section{Summary}
By utilizing large data sets of fiber-aperture and IFU spectroscopic data of low-redshift emission-line galaxies taken from the SDSS DR8 and MaNGA surveys, we investigated the physical and dynamical properties of ionized gas clouds. We summarize the main results as follows.
\begin{enumerate}
  \item The SDSS spectroscopic data suggest that Seyfert galaxies are characterized by a higher electron density and [O~{\sc iii}] velocity dispersion than star-forming galaxies, while LINERs and composite galaxies show intermediate properties between Seyfert galaxies and star-forming galaxies.
  \item The electron density and $\rm{[O\ {\scriptstyle III}]}$ velocity dispersion in NLRs of Seyfert galaxies are positively correlated with the $\rm{[O\ {\scriptstyle III}]\lambda5007/H\beta}$ flux ratio, suggesting that powerful AGN activity causes the enhanced electron density and velocity dispersion.
  \item The $\rm{[O\ {\scriptstyle III}]}$ velocity dispersion in both central and off-central regions is higher than that of star-forming galaxies at a fixed stellar mass.
  \item In the off-central region of Seyfert galaxies, a higher electron density is seen in Seyfert galaxies with an enhanced velocity dispersion than in those without an enhanced velocity dispersion.
\end{enumerate}
All of these results support the idea that the gas clouds in NLRs of Seyfert galaxies come from the nucleus plausibly through the AGN outflow, as predicted by theoretical models such as the radiation-driven fountain model.

\bigskip
\begin{ack}
We thank the anonymous referee, whose comments were very useful to improve this paper.
We also thank S. Koyama, A. Noboriguchi, and N. Tamada, for their fruitful comments. This research is financially supported by the Japan Society for the Promotion of Science (JSPS) KAKENHI 16H03958, 17H01114, 19H00697, and 20H01949. This research made use of ASTROPY, a community-developed core Python package for astronomy.\\
Funding for the Sloan Digital Sky Survey III (SDSS-III) and IV (SDSS-IV, in which the MaNGA survey was carried out) has been provided by the Alfred P. Sloan Foundation, the U.S. Department of Energy Office of Science, the National Science Foundation, and the Participating Institutions. SDSS-IV acknowledges support and resources from the Center for High-Performance Computing at the University of Utah. The SDSS web site is www.sdss.org.
SDSS-III and SDSS-IV are managed by the Astrophysical Research Consortium for the Participating Institutions of the SDSS Collaboration including University of Arizona, the Brazilian Participation Group, Brookhaven National Laboratory, the Carnegie Institution for Science, Carnegie Mellon University, the Chilean Participation Group, University of Florida, the French Participation Group, the German Participation Group, Harvard-Smithsonian Center for Astrophysics, Instituto de Astrof\'isica de Canarias, the Michigan State/Notre Dame/JINA Participation Group, The Johns Hopkins University, Kavli Institute for the Physics and Mathematics of the Universe (IPMU) / University of Tokyo, the Korean Participation Group, Lawrence Berkeley National Laboratory, Leibniz Institut f\"ur Astrophysik Potsdam (AIP), Max-Planck-Institut f\"ur Astronomie (MPIA Heidelberg), Max-Planck-Institut f\"ur Astrophysik (MPA Garching), Max-Planck-Institut f\"ur Extraterrestrische Physik (MPE), National Astronomical Observatories of China, New Mexico State University, New York University, University of Notre Dame, Observat\'ario Nacional / MCTI, The Ohio State University, Pennsylvania State University, University of Portsmouth, Princeton University, Shanghai Astronomical Observatory, the Spanish Participation Group, United Kingdom Participation Group, Universidad Nacional Aut\'onoma de M\'exico, University of Arizona, University of Colorado Boulder, University of Oxford, University of Portsmouth, University of Utah, Vanderbilt University, University of Virginia, University of Washington, University of Wisconsin, Vanderbilt University, and Yale University. 
\end{ack}

\end{document}